\documentstyle[12pt]{article}
\begin{document}
\begin{center}
{\Large \bf New Developments in Cosmology}
\vskip 0.3 true in
{\large J. W. Moffat}
\vskip 0.3 true in
{\it Department of Physics, University of Toronto,
Toronto, Ontario M5S 1A7, Canada}
\end{center}
\vskip 0.3 true in
Talk given at the QCD2000 Workshop, Villefranche-sur-Mer, France,
January, 2000. To be published in the proceedings.
\begin{abstract}%
A brief review is given of the present observational data in
cosmology. A review of a new bimetric gravity theory with
multiple light cones is presented. The physical consequences of
this gravity theory for the early universe are analyzed.
\end{abstract}

\section{\bf Status of Observational Cosmology}

The ten most significant parameters to be determined in
observational cosmology are:
\begin{enumerate}

\item Age of the universe: $t_0$

\item The Hubble constant $H_0$ from the Hubble relation:
$v=H_0d$

\item Density parameter :
$\Omega_m=\frac{\rho_m}{3H^2_0/8\pi G}$

\item Deceleration parameter: $q_0=-\frac{{\ddot R}(t_0)R(t_0)}
{{\dot R}^2(t_0)}$

\item The baryon density $\Omega_B$ and the vacuum density
$\Omega_{\Lambda}$

\item The parameters associated with microwave
backgound fluctuations: $n, \sigma_8, T/S, N_T$

\end{enumerate}

The strongest lower limit for $t_0$ is determined from studies
of the stellar populations of globular clusters. The main error
in the globular clusters age estimate comes from the uncertain
distance to the globular clusters. A 0.25 magnitude error in the
distance translates into a $22\%$ error in the cluster
age~\cite{Chaboyer}. Independent age limits come from the
cooling of white dwarfs. The best estimates give $t_0\sim 13$
Gyr, with a lower limit of $\sim 11$ Gyr. For $t_0 > 13$ Gyr,
we have $h\leq 0.50$ (where $h$ is defined by the Hubble
parameter: $H_0=100$ h km $s^{-1}\, {\rm Mpc}^{-1}$) for matter
density $\Omega_m=1$, and for $h\leq 0.73$ we have $\Omega_m\sim
0.3$ in spatially flat cosmologies with
$\Omega_m+\Omega_{\Lambda}=1$.

The Hubble parameter is now better determined (it used to be
known to within a factor of two). Most measurements are
now consistent with a value: $h=0.65\pm 0.08$~\cite{Saha}. It is
remarkable that data obtained from several different methods for
determing $H_0$ lead to similar results, which gives hope for an
ultimate convergence of measurements. For $\Omega_m=0.4$ and
$\Omega_{\Lambda}=0.6$ and $h=0.65\pm 0.08$, the age of the
universe would be $t_0=13\pm 2$ Gyr, in agreement with globular
cluster age estimates. This result is one of the strong
arguments for a low matter density $\Omega_m\sim 0.3$ and a
non-zero cosmological constant $\Omega_{\Lambda}\sim 0.7$.

A most promising new way of measuring $\Omega_m$ and
$\Omega_{\Lambda}$ on cosmological scales is to use small-angle
anisotropies in the CMB radiation and high-redshift Type Ia
supernovae (SNe Ia). The Supernovae Cosmology
Project (CSP) and the High-Z
Supernovae team have found a significant number of Type Ia
supernovae~\cite{Perlmutter,Garnavich,Riess}. The more recent
larger SCP data set of 42 high redshift data gives for the flat 
case $\Omega_m=0.28^{+0.09+0.05}_{-0.08-0.04}$~\cite{Perlmutter}.
The High-Z Supernovae group has also measured $\Omega_m$ giving 
in the flat case $\Omega_m=0.4\pm 0.3$. Two possible sources of
problems are the dimming by dust and the assumption made that
evolution for nearby and far supernovae is uniform.

In CMB anisotropy studies, the location of the first
acoustic Doppler peak at angular wave number $l\sim 250$ is a
strong indication of a flat universe $\Omega
_m+\Omega_{\Lambda}=1$. The MAXIMA and BOOMERANG balloon flights
seem to confirm this result, and the existence of a second and
possible third peak would appear to be consistent with the
predictions of simple inflation models. New data from the NASDA
Microwave Anisotropy Probe satellite will hopefully strengthen
these results.

We can summarise the main observational results:

\begin{enumerate}

\item   Age of universe$\quad\quad t_0\quad\quad =9-16\,{\rm
Gyr}$ (from globular clusters)
$=9-17\, {\rm Gyr}^*$

\item   Hubble parameter$\quad\quad  H_0\quad \quad
=100\,h\,s^{-1}\,Mpc^{-1},\,\, h=0.65\pm 0.08$.
 
\item   Baryon density$\quad\quad \Omega_bh^2\quad\quad =0.019\pm
0.001$ (from $D/H$)
\vskip 0.1 in
$> 0.015$ from Ly$\alpha$ forest opacity*

\item   Matter density$\quad\quad \Omega_m\quad\quad =0.4\pm 0.2$
(from cluster baryons)
\vskip 0.1 in
$=0.34\pm 0.1$ from Ly$\alpha$ forest $P(k)^*\,(P(k)=Ak^n
\,{\rm with}\,n=1$ for the Harrison-Zel'dovich spectrum)
\vskip 0.1 in
$=0.4\pm 0.2$ from cluster evolution*
\vskip 0.1 in
$> \frac{3}{4}\Omega_{\Lambda}-\frac{1}{4}\pm \frac{1}{8}$ from
SN Ia $> 0.3\,(2.4\sigma$ from flows)

\item   Total density$\quad\quad \Omega_m+\Omega_{\Lambda}=
1\pm 0.3$ (from CMB peak location)

\item   Dark vacuum energy density$\quad\quad \Omega_{\Lambda}= 
0.8\pm 0.3$ (from last two lines)

\item   Neutrino density$\quad\quad \Omega_{\nu}\quad\quad \geq
0.001$ (from Superkamiokande) $\leq 0.1^*$

\end{enumerate}

Here, the cosmological parameters with * are obtained by
assuming $\Lambda CDM$ i.e cold dark matter models with non-zero
cosmological constant.

The Type Ia reshift measurements have indicated the remarkable
result that the cosmic expansion is presently undergoing an
acceleration.

\section {\bf Bimetric Gravity Theory}

A new kind of vector-tensor and scalar-tensor
theory of gravity, which exhibits a bimetric structure and
contains two or more light
cones~\cite{Clayton+Moffat:1999,Clayton+Moffat:1999a,Clayton+Moffat:2000},
has been introduced, recently. This type of model has attracted
some
attention~\cite{Bassett+:2000,Liberati+:2000,Avelino+Martins:2000},
and similar effects have been noted
elsewhere~\cite{Herdeiro:2000,Kiritsis,Brandenberger,Alexander}. 
The motivation for considering these models is derived form 
earlier work ~\cite{Moffat:1993a}, which provided a scenario in 
which some of the outstanding issues in cosmology can be 
resolved. These models provide a fundamental dynamical mechanism 
for varying speed of light theories and generate a new mechanism 
for an inflationary epoch that could solve the initial value 
problems of early universe cosmology. In the following, we shall 
review some of the main features of this new kind of gravity 
theory and its application to cosmology.

In these models matter
that satisfies the strong energy condition can nevertheless
contribute to the cosmic acceleration. Our cosmological model can
be mapped to a model with varying fundamental
constants~\cite{Albrecht+Maguiejo:1999,Barrow+Maguiejo:1998,Barrow:1999},
albeit not uniquely and requiring some care in the interpretation
of the varying constants that appear.

It is hoped that the models can shed some light on the new
observational data that suggests the expansion of the universe at
present is undergoing an
acceleration~\cite{Perlmutter,Garnavich,Bachall:1999}.
Although there has been some success in understanding the latter
problem by the inclusion of a class of very particular scalar
field potentials~\cite{Caldwell}, it is fair
to say that not all issues have been resolved. Using the scalar
field version of the model, we expect that not only will we be
able to generate sufficient inflation, but that a
quintessence-like solution should be achieveable.

We shall be considering models wth an action of the form
\begin{equation}\label{eq:init}
S=\bar{S}_{\rm{gr}}[\bar{g}]+ S[g,\psi]+
\hat{S}[\hat{g},\hat{\phi}^I].
\end{equation}
The first term is the usual Einstein-Hilbert action for general
relativity constructed from a metric $\bar{g}_{\mu\nu}$, and
the final term is the contribution from the non-gravitational
(matter) fields in spacetime $\hat{\phi}^I$, and is built from a
different but related metric $\hat{g}_{\mu\nu}$.

The contribution $S[g,\psi]$ is constructed from a metric
$g_{\mu\nu}$ and includes kinetic terms for a field or
fields (unspecified as yet) $\psi$ that may be considered to be
part of the gravitational sector, modifying the reaction of
spacetime to the presence of the matter fields in
$\hat{S}[\hat{g},\hat{\phi}^I]$. The manner in which
$\psi$ accomplishes this is by modifying the metric that appears
in each of the actions.  For example,
in~\cite{Clayton+Moffat:1999} $\psi$ was a vector field,
$\bar{g}_{\mu\nu}=g_{\mu\nu}$ and
$\hat{g}_{\mu\nu}=g_{\mu\nu}+b\psi_\mu\psi_\nu$,
whereas in~\cite{Clayton+Moffat:1999a} $\psi$ was a scalar field,
$\bar{g}_{\mu\nu}=g_{\mu\nu}$ and
$\hat{g}_{\mu\nu}=g_{\mu\nu}+b\partial_\mu\psi\partial_\nu\psi$.
These relations imply that matter and gravitational fields
propagate at different velocities if $\psi$ is non-vanishing.

Since the matter action $\hat{S}$ is built using only
$\hat{g}_{\mu\nu}$, it is the null surfaces of
$\hat{g}_{\mu\nu}$ along which matter fields propagate. If
we assume that other than the presence of a ``composite'' metric
the matter action is otherwise a conventional form (perfect
fluid, scalar field, Maxwell, {\it etc.}), then variation of
the matter action
yields the matter energy-momentum tensor $\hat{T}^{\mu\nu}$,
which will be conserved
\begin{equation}\label{eq:matter conserve}
\hat{\nabla}_\nu\hat{T}^{\mu\nu}=0,
\end{equation}
by virtue of the matter field equations $\hat{F}_I=0$. Throughout
we will write, for example, $\hat{\nabla}_\nu$ for the covariant
derivative constructed from the Levi-Civita connection of
$\hat{g}_{\mu\nu}$.  Since we also assume that the matter
fields satisfy the dominant energy condition, we therefore know
(assuming appropriate smoothness of $\hat{g}_{\mu\nu}$) that
matter fields cannot travel faster than the speed of light as
determined by $\hat{g}_{\mu\nu}$.

The gravitational action is written
\begin{equation}\label{eq:EH action}
\bar{S}_{\rm{gr}}[\bar{g}]=-\frac{1}{\kappa}\int
d\bar{\mu}\,\bar{R},
\end{equation}
where we use a metric with ($+$$-$$-$$-$) signature and have
defined $\kappa=16\pi G/c^4$.  We will denote the metric
densities by, e.g.,
$\bar{\mu}=\sqrt{-\det(\bar{g}_{\mu\nu})}$ and in addition
write $d\bar{\mu}=\bar{\mu}\,dt\,d^3x$.
We will not consider a cosmological constant, since it can
easily be included later. We can identify the metric
$\bar{g}_{\mu\nu}$ as providing the light cone for the
gravitational system.

We consider a Proca model with arbitrary potential
\begin{equation}
\label{eq:Proca action}
S[g,\psi]=-\frac{1}{\kappa}\int d\mu\Bigl(\frac{1}{4} B^2 -
V(X)\Bigr),
\end{equation}
where we will use the definition
\begin{equation}\label{eq:x}
X=\frac{1}{2}\psi^2,
\end{equation}
and $V^\prime(X)=\partial V(X)/\partial X$.  We will also use
$B_{\mu\nu}=\partial_\mu\psi_\nu-\partial_\nu\psi_\mu$,
$\psi^2=g^{\mu\nu}\psi_\mu\psi_\nu$ and
$B^2=g^{\alpha\mu}g^{\beta\nu}B_{\alpha\beta}B_{\mu\nu}$.
We will assume that as $\psi_\mu\rightarrow 0$ we have $V(X)\sim
m^2X$ and therefore the linearized (in $\psi_\mu$) limit of our
model is identical to Einstein-Proca field equations coupled to
matter. The standard energy-momentum tensor for the vector field
is
\begin{equation}
\label{eq:vector SE}
 T^{\mu\nu}=
 -B^{\mu\alpha}{B^\nu}_\alpha +\frac{1}{4}g^{\mu\nu}B^2
 +V^\prime\psi^\mu\psi^\nu -Vg^{\mu\nu}.
\end{equation}

Although there exists a more general class of models, we will
limit ourselves to the choice
\begin{equation}
\label{eq:metric relation}
\hat{g}_{\mu\nu}=g_{\mu\nu}+b\psi_\mu\psi_\nu,\quad
\bar{g}_{\mu\nu}=g_{\mu\nu}+g\psi_\mu\psi_\nu,
\end{equation}
where $b$ and $g$ are constants, so that the variations of
$\hat{g}_{\mu\nu}$ and $\bar{g}_{\mu\nu}$ are related to those of
$g_{\mu\nu}$ and $\psi_\mu$.

The field equations are given by
\begin{equation}
\label{eq:cov feq}
\label{eq:Proca}
\nabla_\mu B^{\mu\nu}+V^\prime\psi^\nu +g
T^{\mu\nu}\psi_\mu +\kappa\frac{\hat{\mu}}{\mu}(g-b)
\hat{T}^{\nu\mu}\psi_\mu =0,
\end{equation}
\begin{equation}
\bar{\mu}\bar{G}^{\mu\nu} = \frac{1}{2}\mu T^{\mu\nu}
+\frac{1}{2}\kappa\hat{\mu}\hat{T}^{\mu\nu}.
\end{equation}
It is clear that $\hat{g}_{\mu\nu}$ and
$\bar{g}_{\mu\nu}$ provide the characteristic surfaces for
matter and gravitational fields, respectively.

We can prove that any matter model that
conserves energy-momentum with respect to
$\hat{g}_{\mu\nu}$ is consistent with the gravitational
structure that we have introduced~\cite{Clayton+Moffat:2000}.

The ``most physical'' metric is clearly $\hat{g}_{\mu\nu}$,
since it describes the geometry on which matter propagates and
interacts. Because all matter
fields are coupled to the same metric $\hat{g}_{\mu\nu}$ in
exactly the same way, the weak equivalence principle is satisfied.
Furthermore, because one can work in a local Lorentz frame of
$\hat{g}_{\mu\nu}$, in which non-gravitational physics
takes on its special relativistic form, the Einstein equivalence
principle is also satisfied. However, because
$\hat{g}_{\mu\nu}$ does not couple to matter in
the same way as in general relativity unless $\psi_\mu=0$, the
strong equivalence principle will be violated.

The main motivation for considering these theories is that they
should have something to say about the horizon problem in the
early universe. If $\psi_\mu\neq 0$, then if we choose $b>g$,
matter fields will propagate outside the light cone of the
gravitational field.
As $\psi_\mu\rightarrow 0$ the matter light cone will `contract'
and matter and gravitational disturbances will eventually
propagate at the same velocity.  If one considers a frame in which
gravitational waves propagate at a constant speed, then as the
light cone of matter contracts, the universe will appear to
material observers to expand acausally.
 
\section{\bf Cosmology}

Implicit in the idea of a varying light speed is that the speed
of light is changing with respect to some fixed frame of
reference. If one introduces a fundamental frame for this, then
it is perhaps sensible to introduce a function $c(t,x)$ to
describe this
variability~\cite{Albrecht+Maguiejo:1999,Barrow:1999}. The models
that we are considering are based on the idea that the speed of
light can be changing with respect to the speed of gravitational
disturbances, and therefore any indication of the speed of light
as a function of spacetime is frame-dependent. In particular, we
will see that a frame in which the speed of light is constant and
the speed of gravitational disturbances is changing is connected
via a diffeomorphism to a frame where the speed of gravitational
disturbances is constant, and the speed of light is changing.
Quantities of interest such as the local light cone, horizons,
etc. are derived directly from the relevant metric, thereby
avoiding any guesswork as to which `speed of light' to use---the
gravitational or
electromagnetic~\cite{Bassett+:2000,Liberati+:2000}. The constant
$c$ is fixed in the present universe by making measurements of
the electromagnetic field.

In a homogeneous and isotropic (FRW) universe, the vector field
$\psi_\mu$ has components $\psi_\mu=(c\psi_0(\tau),0,0,0)$.  We
will begin with the metric $g_{\mu\nu}$ in comoving form
\begin{equation}
\label{eq:FRWg}
g_{\mu\nu} dx^\mu\otimes dx^\nu =c^2 d\tau\otimes d\tau -
R^2(\tau)\gamma_{ij}dx^i\otimes dx^j,
\end{equation}
and therefore
\begin{equation}
\label{eq:FRWghat}
\hat{g}_{\mu\nu} dx^\mu\otimes dx^\nu
=\hat{\Theta}^2(\tau)c^2 d\tau\otimes d\tau -
R^2(\tau)\gamma_{ij}dx^i\otimes dx^j,
$$ $$
\label{eq:FRWgbar}
 \bar{g}_{\mu\nu} dx^\mu\otimes dx^\nu
=\bar{\Theta}^2(\tau)c^2 d\tau\otimes d\tau
- R^2(\tau)\gamma_{ij}dx^i\otimes dx^j.
\end{equation}
The spatial metric in spherical coordinates has the standard form
\begin{equation}
\gamma_{ij}=\rm{diag}(1/(1-kr^2), r^2, r^2\sin^2\theta),
\end{equation}
and we have defined
\begin{equation}
\hat{\Theta}=\sqrt{1+2bX},\quad{\rm and}\quad
\bar{\Theta}=\sqrt{1+2gX},
\end{equation}
where from~(\ref{eq:x}) we have $X=\frac{1}{2}\psi_0^2$.

Although we begin with the choice~(\ref{eq:FRWg}), once we have
derived the field equations, we will make a coordinate
transformation in order to put $\hat{g}_{\mu\nu}$ in comoving
form and thereby make a comparison with the standard cosmological
results a simpler matter.  Note that we are reversing the
definitions of $t$ and $\tau$ as used in our previous
article~\cite{Clayton+Moffat:1999}.

The matter energy-momentum tensor will have a perfect fluid form
\begin{equation}
 \hat{T}^{\mu\nu}=\Bigl(\rho+\frac{p}{c^2}\Bigr)\hat{u}^\mu\hat{u}^\nu
 -p\hat{g}^{\mu\nu},
\end{equation}
where we have written the velocity field as $\hat{u}^\mu$ to
emphasize that it is normalized using the metric
$\hat{g}_{\mu\nu}$, so that
\begin{equation}
\label{eq:normalization}
\hat{g}_{\mu\nu}\hat{u}^\mu\hat{u}^\nu = c^2.
\end{equation}

The matter conservation laws~(\ref{eq:matter conserve}) lead
to the usual relation
\begin{equation}\label{eq:matter conservation}
\partial_\tau{\rho}+3\frac{\partial_\tau{R}}{R}\Bigl(\rho+\frac{p}{c^2}\Bigr)=0.
\end{equation}
The Friedmann equations take the form
\begin{equation}
\label{eq:psi frame}
\Bigl(\frac{\partial_\tau{R}}{R}\Bigr)^2
 +\frac{kc^2\bar{\Theta}^2}{R^2}=\frac{\kappa c^4}{6}\bar{\Theta}^3
 \Bigl[\frac{\rho}{\hat{\Theta}}+\frac{1}{\kappa c^2}(2XV^\prime-V)\Bigr],
\end{equation}
\begin{equation}
 2\frac{\partial_\tau^2{R}}{R}
 +\Bigl(\frac{\partial_\tau{R}}{R}\Bigr)^2
 +\frac{kc^2\bar{\Theta}^2}{R^2}
 -2\frac{\partial_\tau{R}}{R}\frac{\partial_\tau{\bar{\Theta}}}
 {\bar{\Theta}}
 =-\frac{\kappa c^2}{2}\bar{\Theta}
 \Bigl[\hat{\Theta}p+\frac{1}{\kappa}V\Bigr].
\end{equation}
The single remaining Proca field equation from~(\ref{eq:Proca})
is \begin{equation}
\label{eq:psi feq}
\frac{1}{c\hat{\Theta}}\psi_0
\Bigl[\hat{\Theta}\bigl(\bar{\Theta}^2
V^\prime-gV\bigr)-\kappa(b-g)c^2\rho\Bigr]=0. \end{equation}

We now perform the coordinate transformation
\begin{equation}
dt =\hat{\Theta}d\tau,
\end{equation}
and defining
\begin{equation}
\label{eq:eta definition}
\eta=\frac{\bar{\Theta}}{\hat{\Theta}}=\sqrt{\frac{1+2gX}{1+2bX}},
\end{equation}
we see that the metric $\hat{g}_{\mu\nu}$ is
put into comoving form
\begin{equation}
\label{eq:FRWghat comoving}
 \hat{g}_{\mu\nu} dx^\mu\otimes
 dx^\nu=c^2dt\otimes dt - R^2(t)\gamma_{ij}dx^i\otimes dx^j,
 $$ $$
\label{eq:FRWgbar comoving}
 \bar{g}_{\mu\nu}
 dx^\mu\otimes dx^\nu =\eta^2(t) c^2dt\otimes dt -
 R^2(t)\gamma_{ij}dx^i\otimes dx^j.
\end{equation}
We have
\begin{equation}
\label{eq:matter frame}
\label{eq:physical Friedmann}
\Bigl(\frac{\dot{R}}{R}\Bigr)^2
 +\frac{kc^2\eta^2}{R^2}=\eta^2\frac{\kappa
 c^4}{6}\rho_{\rm{eff}},
\end{equation}
\begin{equation}
  2\frac{\ddot{R}}{R}
 +\Bigl(\frac{\dot{R}}{R}\Bigr)^2 +\frac{kc^2\eta^2}{R^2}
 -2\frac{\dot{R}}{R}\frac{\dot{\eta}}{\eta}= -\eta^2\frac{\kappa
 c^2}{2}p_{\rm{eff}},
\end{equation}
where $\dot{\rho}=\partial_t\rho$.
In~(\ref{eq:matter frame}), we have defined the effective energy
and pressure densities as
\begin{equation}
\label{eq:effectives}
\rho_{\rm{eff}}=\eta\Bigl(\rho+\frac{1}{\kappa
c^2}\hat{\Theta}(2XV^\prime-V)\Bigr),\quad %
p_{\rm{eff}}=\frac{1}{\eta}\Bigl(p+\frac{1}{\kappa\hat{\Theta}}V\Bigr).
\end{equation}
The reason for making these definitions is
that~(\ref{eq:physical Friedmann}) has exactly the form of the
Friedmann equations for the metric $\bar{g}_{\mu\nu}$, and therefore these effective
energy and momentum densities will also satisfy the conservation
laws~(\ref{eq:matter conservation}).

The function $R(t)$ is written in comoving
coordinates and, therefore, the speed of light is constant.  This
emphasizes that having a `varying speed of light' is a
frame-dependent statement.  In a frame where the speed of matter
propagation (including electromagnetic fields) is constant, the
speed of gravitational waves will be changing.  In a frame where
the speed of gravitational waves is constant, the speed of matter
propagation will be changing. This, of course, is as it should
be, since we have not introduced any nondynamical preferred frame
into our model.

In the following we will specialize to a model where the vector
field potential is a simple mass term:
\begin{equation}
\label{eq:potential choice}
V=m^2X,\quad V^\prime =m^2.
\end{equation}
In this case~(\ref{eq:effectives}) becomes
\begin{equation}
\label{eq:effective 2}
\rho_{\rm{eff}}=\eta\Bigl(\rho+(b-g)\rho_{pt}
\hat{\Theta}X\Bigr),\quad
$$ $$
p_{\rm{eff}}
=\frac{1}{\eta}\Bigl(p+c^2(b-g)\rho_{pt}
\frac{X}{\hat{\Theta}}\Bigr).
\end{equation}

The nontrivial solution ($\psi_0\neq 0$) of the field
equation~(\ref{eq:psi feq}) leads to
\begin{equation}
\label{eq:spec psi}
\rho=\rho_{pt}\hat{\Theta}(1+gX),
\end{equation}
where
\begin{equation}
\rho_{pt}=\frac{m^2}{\kappa c^2(b-g)},\quad
H_{pt}=\sqrt{\frac{c^2m^2}{6(b-g)}},
\end{equation}
are the density at which $\psi_0^2=0$ is reached, and the inverse
Hubble time at which this occurs (assuming that $k=0$).

We can now write the acceleration parameter as observed by
material observers from~(\ref{eq:matter frame}) as
\begin{equation}
\label{eq:q}
 \hat{q}=-\frac{\ddot{R}}{H^2R}=\frac{\kappa c^4}{12}\frac{\eta^2}{H^2}
 \Bigl(\rho_{\rm{eff}}+\frac{3}{c^2}p_{\rm{eff}}\Bigr)
 -\frac{\dot{\eta}}{H\eta},
\end{equation}
where we have defined the Hubble function $H=\dot{R}/R$. We have
\begin{equation}
\label{eq:eta dot}
\frac{\dot{\eta}}{H\eta}
 =3\frac{(b-g)}{\rho_{pt}\bar{\Theta}^2\hat{\Theta}(g+b+3bgX)}\Bigl(\rho+\frac{p}{c^2}\Bigr).
\end{equation}

\section{\bf The Very Early Universe}

For very short times following the initial singularity, we expect
that $\psi_0$ is large, and if we assume that $gX\gg 1$ and $bX\gg
1$, then from~(\ref{eq:spec psi}) we find that
\begin{equation}\label{eq:spec rho}
\rho=\rho_{pt}\sqrt{2b}gX^{3/2}.
\end{equation}
This results in the Friedmann equation
\begin{equation}
\label{eq:early Friedmann}
 \Bigl(\frac{\dot{R}}{R}\Bigr)^2
 +\frac{k\bar{c}^2}{R^2}=\frac{\bar{\kappa} \bar{c}^4}{6}\rho,
\end{equation}
where
\begin{equation}
 \bar{c}=c\sqrt{\frac{g}{b}},\quad
 \bar{G}=G\sqrt{\frac{g}{b}},\quad
 \bar{\kappa}=\frac{16\pi \bar{G}}{\bar{c}^4}.
\end{equation}

Although the behaviour of the solutions 
are well-known, it is worth
pointing out that the `effective' constants $\bar{c}$ and
$\bar{G}$ are not interpretable as the effective speed
of light and gravitational constant, rather they are effective
constants that dictate how the gravitational field reacts to the
presence of matter. Matter fields continue to propagate with
speed $c$ consistent with~(\ref{eq:FRWghat comoving}). It is the
gravitational field perturbations that propagate with speed
$\bar{c}$, which is the justification for the notation.

During this phase there is clearly no inflation, but the horizon
scales of the gravitational field and matter fields are related
by
\begin{equation}
\bar{d}_H(t)=\frac{\bar{c}}{c}\hat{d}_H(t)=\sqrt{\frac{g}{b}}
\hat{d}_H(t),\quad{\rm
where}\quad \hat{d}_H(t)=cR(t)\int_0^t\frac{ds}{R(s)},
\end{equation}
with a similar definition for $\bar{d}_H(t)$ using
the metric ${\bar g}_{\mu\nu}$. Because we have $g<b$ we
expect that not only is the speed of gravitational disturbances
slower than that of matter, but also that the coupling between
matter and the gravitational sector is also lessened.

What we have here is very close to what was originally envisaged
by one of us in~\cite{Moffat:1993a}.
This is part of the motivation for including the $g\ne 0$
possibility, the other is that the approach to the initial
singularity in this phase follows the same path as in ordinary
GR+matter models, with a re-interpretation of the parameters. In
this case we have a model that interpolates between this initial
period where $\bar{c}>c$ and the later universe where
$\bar{c}=c$.

\section{\bf Inflation and Light Cone Contraction}

As $\psi_0$ decreases towards the point where $gX\sim 1$ the
solution will no longer be a
good approximation.  If we now consider the solution when $gX\ll
1$, from~(\ref{eq:spec psi}) we have
\begin{equation}\label{eq:late rho x}
 \hat{\Theta}=\frac{\rho}{\rho_{pt}},\quad {\rm or}\quad
 X=\frac{1}{2b}\Bigl[\Bigl(\frac{\rho}{\rho_{pt}}\Bigr)^2-1\Bigr],
\end{equation}
and the Friedmann equation~(\ref{eq:physical Friedmann}) becomes
\begin{equation}
 \Bigl(\frac{\dot{R}}{R}\Bigr)^2
 +\frac{kc^2\eta^2}{R^2}\Bigl(\frac{\rho_{pt}}{\rho}\Bigr)^2
 =\frac{\kappa
 c^4}{12}\rho_{pt}\Bigl[1+\Bigl(\frac{\rho_{pt}}
 {\rho}\Bigr)^2\Bigr].
\end{equation}

In this limit
\begin{equation}
 \rho_{\rm{eff}}+\frac{3}{c^2}p_{\rm{eff}}
 =\frac{1}{\rho_{pt}}\Bigl[
 \rho\Bigl(\rho+\frac{3}{c^2}p\Bigr)+\rho^2-\rho^2_{pt}\Bigr],
\end{equation}
which is greater than zero if the strong energy condition is
satisfied, since $\rho\ge \rho_{pt}$, and~(\ref{eq:q})
reduces to
\begin{equation}\label{eq:q hat}
 \hat{q}=\frac{\kappa c^4\rho_{pt}}{12H^2}
 \Bigl[\frac{1}{\rho}\Bigl(\rho+\frac{3}{c^2}p\Bigr)+1-
 \Bigl(\frac{\rho_{pt}}{\rho}\Bigr)^2\Bigr]
-\frac{3}{\rho}\Bigl(\rho+\frac{p}{c^2}\Bigr).
\end{equation}
Since we expect that $H^2$ is large in the early universe (we can
arrange that $\rho_{pt}\ll \rho_c$ where $\rho_c=12
H^2/(\kappa c^4)$, it is clear from~(\ref{eq:q hat}) that even
if matter satisfies the strong energy condition, the final term
will dominate and $\hat{q}<0$ (unless, perhaps, the weak energy
condition is also violated). This is the expansion of the
universe as seen by material observers. The acceleration of the
gravitational geometry $\bar{q}$ would lack the final term and
therefore $\bar{q}>0$.

That we get inflation was demonstrated
previously~\cite{Clayton+Moffat:1999}, where an exact solution
for $k=0$ and $g=0$ was found.  Although we discovered that we
could not get enough expansion to solve the horizon problem with
pure radiation, a slowly rolling scalar field could provide the
necessary negative pressure.  The role that the extra structure
of our model plays is that the fine-tuning that is required in a
simple scalar field, potential-driven model is alleviated.

The flatness problem requires a bit more explanation.
Dividing~(\ref{eq:physical Friedmann}) by $H^2$ and defining
\begin{equation}\label{eq:epsilon}
\bar{\epsilon}=\frac{kc^2\eta^2}{(\dot{R})^2},
\end{equation}
we find a differential equation that $\bar{\epsilon}$ satisfies by
taking a derivative and using~(\ref{eq:matter frame}) to give
\begin{equation}
\dot{\bar{\epsilon}}=\frac{\kappa
c^4\eta^2}{6H}\bar{\epsilon}\Bigl(\rho_{\rm{eff}}
+\frac{3}{c^2}p_{\rm{eff}}\Bigr).
\end{equation}
Therefore, since $\bar{\epsilon}>0$ and $H>0$ in the early
universe, the only way for $\bar{\epsilon}=0$ to be an attractor
for~(\ref{eq:physical Friedmann}) is for
$\rho_{\rm{eff}}+\frac{3}{c^2}p_{\rm{eff}}<0$ at least
for part of the history of the universe.  What is not so obvious
is whether the quantity $\bar{\epsilon}$ as defined
in~(\ref{eq:epsilon}) is of physical relevance.

The quantity of geometrical importance is the $3$-curvature of
the spacelike slices, $6k/R^2$, which suggests that the
physically meaningful quantity to examine would be
\begin{equation}
\label{eq:epsilon hat}
\hat{\epsilon}=\frac{kc^2}{(\dot{R})^2},
\end{equation}
which has the equation of motion
\begin{equation}\label{eq:epsilon hat dot}
\dot{\hat{\epsilon}}=2\hat{\epsilon}\hat{q}.
\end{equation}
Another way of stating this is that the curvature radius defines
$\Omega$ through
\begin{equation}
R_{\rm curv}=\frac{R}{\vert
k\vert^{1/2}}=\frac{c}{H(\vert\Omega-1\vert)^{1/2}},
\end{equation}
and so $\hat{\epsilon}=\vert\Omega-1\vert$. Since we found
from~(\ref{eq:q hat}) that $\hat{q}<0$ in the early universe,
clearly $\hat{\epsilon}=0$ is an attractor for~(\ref{eq:physical
Friedmann}), and since it is most-likely the quantity of physical
importance for matter physics, we can also claim to have solved
the flatness problem once the horizon problem is solved.

\section{\bf Conclusions}

In our bimetric model, the universe generically accelerates
($\hat{q}<0$) during some period in the early universe, and
in the same period the physical importance of spatial
curvature diminishes ($\vert\Omega-1\vert$ is decreasing).  This
can occur even when the matter fields satisfy the strong energy
condition.

The model that we have considered generalizes that which appeared
in~\cite{Clayton+Moffat:1999,Clayton+Moffat:2000} in a way that
more closely follows the scenario discussed
in~\cite{Moffat:1993a}. In the very early universe, matter and
gravitational fields propagate with different and approximately
constant velocities. During a period in which the matter
light cone, originally much larger than the light cone of
gravity, contracts, material observers will see an acausal
expansion of the universe similar to inflation. Because the light
cone of gravity does not undergo the same contraction, we expect
there to be an observable difference in the scalar versus tensor
contributions to the cosmic microwave background anisotropies.

\section{Acknowledgements}

This work was supported by the Natural Sciences and Engineering
Research Council of Canada.
\vskip 0.3 true in

\end{document}